
\magnification=1200
\baselineskip=10pt
\def\lsim{<\kern-2.5ex\lower0.85ex\hbox{$\sim$}\ }
\def\rsim{>\kern-2.5ex\lower0.85ex\hbox{$\sim$}\ }
\overfullrule=0pt
\def\LAMBDABAR {\hbox{$\lambda$\kern-0.52em\raise+0.45ex\hbox{--}\kern+0.2em}}
\def\ebar {\hbox{E\kern-0.6em\raise0.2ex\hbox{/}\kern+0.1em}}
\rightline{UR--1284\ \ \ \ \ \ \ \ }
\rightline{ER--40685--734}
\baselineskip=20pt
\centerline{\bf SELF-DUALITY AND THE SUPERSYMMETRIC KdV HIERARCHY}
\vskip 2cm
\centerline{Ashok Das}
\centerline{and}
\centerline{C. A. P. Galv\~ao \footnote*{Permanent address:
CBPF, Dept. of Fields and Particles, Rua Dr Xavier Sigaud 150,
Urca, 22290 Rio de Janeiro, Brazil}}
\centerline{Department of Physics and Astronomy}
\centerline{University of Rochester}
\centerline{Rochester, NY 14627, USA}
\vskip 2cm
\centerline{\underbar{Abstract}}
\bigskip
We show how the  supersymmetric KdV equation
 can be obtained from the self--duality
condition on Yang--Mills fields in four dimension associated with the
 graded Lie
algebra OSp(2/1).  We also obtain the hierarchy of Susy KdV equations from
such a condition.  We formulate the Susy KdV hierarchy as a vanishing
curvature condition associated with the U(1) group and
 show how an Abelian
self--duality condition in four dimension can also lead to these equations.
\vfil\eject
\noindent {\bf I. \underbar{Introduction}:}

It is by now known that most of the bosonic integrable equations [1--2]
 in $1+1$
and $2+1$ dimensions can be obtained from a self--duality condition on
Yang--Mills fields in higher dimensions belonging to the Lie algebra
SL(2,{\bf R}) [3--6].  In a recent paper, we showed [7] how this method can be
generalized to obtain the entire hierarchy of the KdV equations along with
the appropriate recursion relation from a self--duality condition in four
dimension.  The extension of these studies to fermionic integrable systems
has not yet been carried out.  The main difficulties are as follows.
Fermionic integrable systems can be either supersymmetric or
nonsupersymmetric.  Furthermore, even for the supersymmetric ones, a zero
curvature formulation in superspace does not yet exist.  The question of
self--duality is, therefore, rather a tricky one.

Instead, the fermionic integrable systems such as the s-KdV [8]
 or the Susy KdV [9]
equations are known to result from a zero curvature formulation associated
with the graded Lie algebra OSp(2/1) [10].  In this letter, we study the
self--duality condition associated with Yang--Mills fields belonging to the
graded Lie algebra OSp(2/1) in four dimensions and show how the Susy KdV
equation as well as the hierarchy of these equations can
 be obtained from
these conditions.  Unfortunately, we have not been able to derive the
s--KdV equation from these self--duality conditions.  We also show how the
hierarchy of Susy KdV equations can be obtained from the vanishing
curvature of an Abelian gauge superfield.  This, in turn, shows that the
hierarchy of equations can also
 be obtained from a self--duality condition on an
Abelian gauge superfield in four dimensions.
\medskip
\noindent {\bf II. \underbar{Self--duality and the Susy KdV Equation}:}

The supersymmetric KdV equation in $1+1$ dimension involves two dynamical
variables, [8--10]
 $u(x,t)$ and $\phi (x,t)$ where $u(x,t)$ is a bosonic variable
while $\phi (x,t)$ is fermionic.  The system of equations have the form
$$\eqalign{{\partial u \over \partial t} &= {1 \over 2}\
\partial^3_x u + 3 u \partial_x u + {3 \over 2}\
\partial^2_x \phi \phi\cr
\noalign{\vskip 4pt}%
{\partial \phi \over \partial t} &= {1 \over 2}\
\partial^3_x \phi + {3 \over 2}\ \partial_x (u \phi)\cr}\eqno(1)$$
where $\partial_x = {\partial \over \partial x}$.  These equations are
invariant under the global supersymmetry transformations
$$\eqalign{\delta u &= \epsilon \partial_x \phi \cr
\delta \phi &= \epsilon u \cr}\eqno(2)$$
where $\epsilon$ is a constant anticommuting parameter (Grassmann
parameter).  The Susy KdV equations (Eq. (1)) are known to be integrable
and can be obtained from a zero curvature condition associated with the
graded Lie algebra of OSp(2/1). Thus identifying
$$\eqalign{A_x(x,t) &= \pmatrix{0 &-u &i \phi\cr
\noalign{\vskip 6pt}%
1 &0 &0\cr
\noalign{\vskip 6pt}%
0 &-i \phi &0\cr}\cr
\noalign{\vskip 10pt}%
A_t(x,t) &= \pmatrix{- {1 \over 2}\ \partial_x u
&- {1 \over 2}\ \partial^2_x u - u^2 -
{1 \over 2}\ \partial_x \phi \phi
&{i \over 2}\ \partial^2_x \phi + i u \phi\cr
\noalign{\vskip 6pt}%
u &{1 \over 2}\ \partial_x u
&- {i \over 2}\ \partial_x \phi\cr
\noalign{\vskip 6pt}%
- {i \over 2}\ \partial_x \phi &- {i \over 2}\ \partial^2_x \phi - i u \phi
&0\cr}\cr}\eqno(3)$$
we note that the vanishing curvature condition
$$\partial_t A_x - \partial_x A_t + [A_t , A_x ] = 0$$
\smallskip
$${\rm or,}\quad \pmatrix{0 &-{\partial u \over \partial t} + {1 \over
 2}\ \partial^3_x u + 3 u \partial_x u + {3 \over 2}\
\partial^2_x \phi \phi &i \big( {\partial \phi \over
\partial t} - {1 \over 2}\ \partial^3_x \phi -
{3 \over 2}\ \partial_x (u \phi) \big)\cr
\noalign{\vskip 6pt}%
0 &0 &0\cr
\noalign{\vskip 6pt}%
0 &-i \big( {\partial \phi \over \partial t} - {1 \over 2}\ \partial^3_x
\phi - {3 \over 2}\ \partial_x (u \phi) \big) &0\cr} = 0 \eqno(4)$$
\smallskip
\noindent gives Eq. (1) or the Susy KdV equation.

We can now follow the standard formulation [5--6]
 of the self--duality condition
in a four dimensional space with (2,2) signature.  By appropriately
choosing the null direction, we can write the self--duality condition as
$$\eqalign{&[\partial_t - H, \partial_x - Q] = 0\cr
&[P,B] = 0\cr
&\partial_x (Q - P) + [Q,P] = [H,B]\cr}\eqno(5)$$
where the four dimensional Yang--Mills potentials, $Q,\ H,\ P\ {\rm and}
\ B$ are assumed to be matrices belonging to OSp(2/1) and depending only on
the coordinates $x \ {\rm and}\ t$.  It is easy to check that the matrices
$$\eqalign{Q &= -A_x = \pmatrix{0 &u &- i \phi\cr
\noalign{\vskip 6pt}%
-1 &0 &0\cr
\noalign{\vskip 6pt}%
0 &i \phi &0\cr}\cr
\noalign{\vskip 10pt}%
H &= -A_t = \pmatrix{{1 \over 2}\ \partial_x u
&{1 \over 2}\ \partial^2_x u + u^2 + {1 \over 2}\ \partial_x \phi \phi
&- {i \over 2}\ \partial^2_x \phi - iu \phi\cr
\noalign{\vskip 6pt}%
-u &- {1 \over 2}\ \partial_x u
&{i \over 2} \ \partial_x \phi\cr
\noalign{\vskip 6pt}%
{i \over 2}\ \partial_x \phi
&{i \over 2}\ \partial^2_x \phi + i u \phi
&0\cr}\cr
\noalign{\vskip 10pt}%
P &= \pmatrix{0 &{1 \over 2}\ u &- {3i \over 4}\ \phi\cr
\noalign{\vskip 6pt}%
0 &0 &0\cr
\noalign{\vskip 6pt}%
0 &{3i \over 4}\ \phi &0\cr}\cr
\noalign{\vskip 10pt}%
B &= \pmatrix{0 &{1 \over 2} &0\cr
\noalign{\vskip 6pt}%
0 &0 &0\cr
\noalign{\vskip 6pt}%
0 &0 &0\cr}\cr}\eqno(6)$$
satisfy the self--duality condition of Eq. (5) and give rise to the Susy KdV
 equation.

We would like to point out that the form of the self--duality condition in
Eq. (5) is not suitable to obtain the hierarchy of Susy KdV equations
 as also is true in the bosonic case [7].  We
have also not been able to obtain the s--KdV equation [8], which is not
supersymmetric, from the self--duality condition of Eq. (5).
\medskip
\noindent {\bf III. \underbar{Self--duality and the Susy KdV Hierarchy}:}

To obtain the hierarchy of Susy KdV equations, we follow the method in ref.
  7.  Thus, we identify the coordinates of the four dimensional space with
(2,2) signature as $x^0 = t,\ x^1 = x,\ x^2 = \tilde t$ and
$x^3 = \tilde x$.  If we choose
$$\epsilon^{0123} = 1 = \epsilon_{0123} \eqno(7)$$
then the self--duality condition takes the form
$$F_{\mu \nu} = {1 \over 2}\ \epsilon_{\mu \nu}^{\ \ \lambda \rho}
F_{\lambda \rho} \qquad\qquad \mu, \nu, \lambda, \rho = 0,1,2,3
\eqno(8)$$
where the field strength, in general for a matrix potential, is defined to
be
$$F_{\mu \nu} = \partial_\mu A_\nu - \partial_\nu A_\mu + [
A_\mu , A_\nu] \eqno(9)$$
Writing out explicitly the self--duality condition in Eq. (8), we obtain
$$F_{01} = - F_{23}$$
$${\rm or,}\qquad \partial_0 A_1 - \partial_1 A_0 + [A_0 , A_1] =
- (\partial_2 A_3 - \partial_3 A_2 + [A_2 , A_3])
\eqno(10)$$
$$F_{02} = - F_{13}$$
$${\rm or,}\qquad \partial_0 A_2 - \partial_2 A_0 + [A_0 , A_2] = -
(\partial_1 A_3 - \partial_3 A_1 + [A_1 , A_3 ])
\eqno(11)$$
$$F_{03} = - F_{12}$$
$${\rm or,}\qquad \partial_0 A_3 - \partial_3 A_0 +
[A_0 , A_3] = -
(\partial_1 A_2 - \partial_2 A_1 +
[A_1 , A_2])\eqno(12)$$

If we now assume the gauge potentials to be independent of the coordinates
$(x^3 , x^4)$ and identify
$$A_2 = A_3 \eqno(13)$$
then the self--duality conditions reduce to
$$\eqalignno{\partial_0 A_1 - \partial_1 A_0 + [A_0 , A_1] &= 0
 &(14)\cr
(\partial_0 + \partial_1)A_2 + [A_0 + A_1, A_2] &= 0 &(15)\cr}$$
Note that if we choose the potentials to be matrices belonging to
OSp(2/1) and identify
$$\eqalign{A_1 &= \pmatrix{0 &-u &i \phi\cr
\noalign{\vskip 6pt}%
1 &0 &0\cr
\noalign{\vskip 6pt}%
0 &-i \phi &0\cr}\cr
\noalign{\vskip 10pt}%
A_0 &= \pmatrix{- {1 \over 2}\ \partial_x C &- {1 \over 2}\
\partial^2_x C - u C + {1 \over 2}\ G \phi
&- {i \over 2}\ \partial_x G + i C \phi\cr
\noalign{\vskip 6pt}%
C &{1 \over 2}\ \partial_x C &{i \over 2}\ G\cr
\noalign{\vskip 6pt}%
{i \over 2}\ G &{i \over 2} \ \partial_x G - i C \phi
 &0\cr}\cr}\eqno(16)$$
then Eq. (14) gives
$$\eqalign{{\partial u \over \partial t} &= {1 \over 2}\ \big( \partial^3_x
+ 2 (\partial_x u + u \partial_x)\big)C - {3 \over 2}\ \partial_x
 G \phi - {1 \over 2}\ G \partial_x \phi\cr
\noalign{\vskip 4pt}%
{\partial \phi \over \partial t} &= - {1 \over 2}\
\partial^2_x G - {1 \over 2}\ u G + {3 \over 2}\
\partial_x C \phi + C \partial_x \phi\cr}\eqno(17)$$
This is the supersymmetric nth order KdV equation where
$$C = {\delta H_n \over \delta u (x,t)} \eqno(18)$$
with $H_n$ the nth conserved charge of the KdV equation and $G$ can be
obtained from the supersymmetric variation of $C$.  The coupled equations
in Eq. (17) are invariant under the global supersymmetry transformations
$$\eqalign{\delta u &= \epsilon \partial_x \phi \cr
\delta \phi &= \epsilon u \cr
\delta C &= - \epsilon G\cr
\delta G &= - \epsilon \partial_x C\cr}\eqno(19)$$
With the choice of $A_0$ and $A_1$ in Eq. (16), we can now solve Eq. (15),
at least formally, as
$$A_2 (x^+ , x^-) = A_3 (x^+ , x^-)
= e^{- \int^{x^+} dx^{\prime +}A_+(x^{\prime +} , x^-)}
A_2 (0 , x^-) e^{\int^{x^+}
dx^{\prime \prime +} A_+ (x^{\prime \prime +} , x^-)}
\eqno(20)$$
where  we have defined
$$\eqalign{x^{\pm} &= x \pm t\cr
A_+ &= A_0 + A_1 \cr}\eqno(21)$$
and $A_2 (0, x^-)$ is any OSp(2/1) matrix
 depending only on the coordinate $x^-$.
This, therefore, shows how the hierarchy of Susy KdV equations can be
obtained from the self--duality condition on Yang--Mills potentials in four
dimensions belonging to the graded Lie algebra OSp(2/1).
\medskip
\noindent {\bf IV. \underbar{Zero Curvature Formulation Associated
 with U(1)}:}

As we have seen, the hierarchy of supersymmetric KdV equations have the
form
$$\eqalign{{\partial u \over \partial t} &= {1 \over 2}\
\big( \partial^3_x + 2 (\partial_x u + u \partial_x) \big) C -
 {3 \over 2}\ \partial_x G \phi - {1 \over 2}\ G
 \partial_x \phi\cr
\noalign{\vskip 4pt}%
{\partial \phi \over \partial t} &= - {1 \over 2}\ \partial^2_x G - {1
 \over 2}\ uG + {3 \over 2}\ \partial_x C \phi + C
 \partial_x \phi\cr}\eqno(22)$$
where $C$ and $G$ represent appropriate bosonic and fermionic functionals
of the dynamical variables $u(x,t)$ and $\phi (x,t)$.  This system, as we
noted, is supersymmetric as well as integrable.  Therefore, there exist an
infinite number of conserved charges associated with this system.  Of
particular interest are the supersymmetry charge defined to be
$$Q = \int dx\ \phi (x,t) \eqno(23)$$
and the lowest order bosonic conserved charge
$$H_0 = \int dx \ u(x,t) \eqno(24)$$
This charge in Eq. (24) can, in fact, be identified with $Q^2$ up to a
multiplicative constant.  From the fact that these charges are conserved,
it follows that
$${\partial H_0 \over \partial t} = \int dx \ {\partial u
\over \partial t} = 0$$
$${\rm or,}\qquad\qquad {\partial u \over \partial t} = {\partial K  \over
\partial x}\eqno(25)$$
$${\partial Q \over \partial t} = \int
dx\ {\partial \phi \over \partial t}$$
$${\rm or,}\qquad\qquad {\partial \phi \over \partial t} = {\partial
 \psi \over
\partial x}\eqno(26)$$
Namely, the right hand sides of Eq. (22) can be identified with total
derivatives as
$$\eqalign{{\partial u \over \partial t} &= {\partial K \over \partial x}
 = {1 \over 2}\ \big( \partial^3_x + 2(\partial_x u +
u \partial_x )\big) C -
{3 \over 2}\ \partial_x G \phi - {1 \over 2}\
G \partial_x \phi\cr
\noalign{\vskip 4pt}%
{\partial \phi \over \partial t} &= {\partial \psi \over \partial x}
 = - {1 \over 2}\  \partial^2_x G - {1 \over 2}\
  uG +
{3 \over 2}\ \partial_x C \phi + C
\partial_x \phi\cr}\eqno(27)$$
It is straightforward to show that for these equations to be invariant
under the supersymmetry transformations of Eq. (19), we must have
$$\eqalign{\delta K &= \epsilon \partial_x \psi\cr
\delta \psi &= \epsilon K\cr}\eqno(28)$$

Let us now define a simple superspace parameterized by the coordinates
$(x, \theta)$ where $\theta$ is the Grassmann coordinate.  On this space,
we can define the supersymmetric charge operator as
$$Q = {\partial \over \partial \theta} + \theta \ {\partial
 \over \partial x} \eqno(29)$$
and would generate the translations
$$\eqalign{\delta_\epsilon \theta &= \epsilon\cr
\delta_\epsilon x &=  \epsilon \theta\cr}\eqno(30)$$
On this space, we can define two fermionic superfields as U(1) gauge
potentials of the form
$$\eqalign{A_x (x, \theta) &= \phi + \theta u\cr
A_t (x, \theta) &= \psi + \theta K\cr}\eqno(31)$$
This guarantees the appropriate transformations of the variables under
supersymmetry as given in Eqs. (19) and (28).  The associated U(1) field
strength, in this case, will be a superfield of the form
$$\eqalign{F_{tx} &= \partial_t A_x (x, \theta) - \partial_x
 A_t (x, \theta)\cr
&= ( \partial_t \phi - \partial_x \psi)
 + \theta (\partial_t u - \partial_x K)\cr}\eqno(32)$$
Consequently, the vanishing of the U(1) curvature superfield, in the
present case, gives back the supersymmetric KdV hierarchy equations as in
Eq. (27).

We would like to point out here that we have studied the vanishing field
strength on the supermanifold as well.  However, unlike the case of the BRS
transformations, which can be obtained from a vanishing curvature on a
supermanifold, here such a condition runs into inconsistencies.
Furthermore, we would like to note that for the s--KdV equations [8] which
are not supersymmetric, there is no analogue of the fermionic charge of
Eq. (23).  Correspondingly, the fermionic equation of the s--KdV equation
cannot be written as
$$\partial_t \phi = \partial_x \psi \eqno(33)$$
where $\psi$ is a local functional of the dynamical variables.
Consequently, an Abelian zero curvature formulation does not exist for the
s--KdV equation in terms of local potentials.

Given the zero curvature formulation based on the group U(1), the
derivation of the Susy KdV hierarchy follows in a straightforward
manner from the self--duality conditions of Eqs. (13--15).
  In this case, if we choose the potentials to be
 Abelian superfields, then the equations take the form
$$\eqalign{\partial_0 A_1 - \partial_1 A_0 &= 0\cr
(\partial_0 + \partial_1) A_2 &= 0\cr}\eqno(34)$$
If we identify
$$\eqalign{A_1 (x, \theta) &= A_x (x, \theta) = \phi + \theta u\cr
A_0 (x, \theta) &= A_t (x, \theta) = \psi
 + \theta K\cr
A_2 (x, \theta) &= A_3 (x, \theta) = A_2
(t - x, \theta) = \ {\rm arbitrary}\cr}\eqno(35)$$
then the self--duality conditions of Eq. (34) are automatically satisfied
and lead to the hierarchy of Susy KdV equations.
\medskip
\noindent {\bf V. \underbar{Conclusion}:}

We have derived the Susy KdV equation as well as the hierarchy of such
equations from the self--duality conditions on Yang--Mills potentials
belonging to the graded Lie algebra OSp(2/1) in four dimensions.  We have
shown how the hierarchy of Susy KdV equations can be formulated as a zero
curvature associated with the U(1) group and have generalized the
self--duality construction to this case as well.

We would like to thank Wen--Jui Huang for many helpful discussions.  This
work was supported in part by the U.S. Department of Energy Grant No.
 DE--FG02--91ER40685 and by CNPq, Brazil.
\bigskip
\bigskip
\noindent {\bf References:}
\item{1.} L. D. Faddeev and L. A. Takhtajan, \lq\lq Hamiltonian methods in
the theory of solitons", Springer (Berlin 1987).
\item{2.} A. Das, \lq\lq Integrable Models", World Scientific
(Singapore 1989).
\item{3.} R. S. Ward, Nucl. Phys. {\bf B236} (1984) 381; Philos. Trans. R.
Soc. {\bf A315} (1985) 451; also in \lq\lq Field theory, quantum gravity
and strings", vol. 2, eds. H. J. de Vega and N.~Sanchez, p106.
\item{4.} L. J. Mason and G. A. J. Sparling, Phys. Lett. {\bf A137}
 (1989) 29.
\item{5.} I. Bakas and D. A. Depireux, Int. J. Mod. Phys.
 {\bf A7} (1992) 1767.
\item{6.} A. Das, Z. Khviengia and E. Sezgin, Phys. Lett.
{\bf B289} (1992) 347.
\item{7.} A. Das and C. A. P. Galv\~ao, University Rochester preprint
 UR--1274.
\item{8.} B. A. Kupershmidt, Phys. Lett. {\bf A102} (1984) 213.
\item{9.} Y. Manin and A. O. Radul, Comm. Math. Phys. {\bf 98} (1985) 65.
\item{  } P. Mathieu, Phys. Lett. {\bf B203} (1988) 287; J.~Math. Phys.
{\bf 29} (1988) 2499.
\item{10.} M. Gurses and O. Oguz, Lett. Math. Phys. {\bf 11} (1986) 235.
\item{   } A. Das and S. Roy, J. Math. Phys. {\bf 31} (1990) 2145.

\end